*Powder Ball Milling: An energy balance approach to particle size reduction*

Stefano Martelli
RINA Consulting - Centro Sviluppo Materiali S.p.A., Rome, ITALY
stefano.martelli55@gmail.com - ORCiD: 0009-0009-5808-9056

Paolo Emilio Di Nunzio*
RINA Consulting - Centro Sviluppo Materiali S.p.A.,
Via di Castel Romano 100, 00128 Rome, ITALY
paolo.dinunzio@rina.org - ORCiD: 0000-0001-8260-6644
* Corresponding author

Abstract
Milling of brittle materials shows a lower limit in size reduction, below which grinding cannot proceed further. Experimentally, the smallest particle size is determined by the milling media energy, and it is time independent. Rittinger's law, which applies to fine grinding, does not consider any lower limit in size, leading to nonphysical results. As a modification, a number of dampening parameters that gradually reduce grinding efficiency have already been proposed. In this work, the grinding inhibition is linked to an energy balance between the powder's specific surface energy and the energy density, that is the kinetic energy of the milling media divided by the impacted volume. Introducing the energy equilibrium concept, the comminution kinetics can be formally described by a JMAK-like equation. Certain parallels are indicated. Validation of this formulation has been conducted on published and original data.

Abbreviations
BPVR = Ball to Powder Volume Ratio
DEM = Discrete Element Method
JMAK = Johnson–Mehl–Avrami–Kolmogorov
LD = Laser Diffraction
MA = Mechanical Alloying
PSD = Particle Size Distribution
BET = Brunauer-Emmett-Teller

1. Introduction

The all-important size reduction process, also known as comminution, is defined as the mechanical breakdown of solids into smaller particles without changing their state of aggregation [1]. The ball mill is one of various size-reduction tools. It is a crushing machine with a revolving drum that holds ceramic or metal balls or pebbles for grinding purposes. Ball milling is a development of the mortar and pestle, which was created in the Stone Age and has mostly not changed over time. It is extensively used for manufacturing cement, paints, pyrotechnics, ceramics. More recently, ball milling has assisted the progress of ultra-fine and nano-sized reactive powder, as well as the foundation of new disciplines such as mechanical-alloying [2] and mechano-chemistry [3].

The realization of sub-micron and nano-sized metal and ceramic powders led to the design of high energy grinding units, such as the Szegvari attrition mill [4], firstly used to create fine sulfur dispersions for rubber vulcanization, and the high energy planetary ball mill [5], introduced by Fritsch GmbH in 1961. Grinding down to the ultra-fine and nano-size dimension asks for ball impact energy above the values accessible to the conventional horizontal tumbler ball mill, where the maximum impact energy, in cataracting regime [6], is essentially governed by the height reached by the milling media vessel during turning before dropping by gravity to accomplish the grinding action. Independently on their design and configuration, all the comminution equipment shows some common features:

Very little of the power [7,8], primarily electrical, used by the milling devices is actually converted to grinding; the majority of the power is lost as heat. High energy ball milling apparatuses need strong cooling in order to run for extended periods of time.

The reduction in particle size is not proportional to milling time. It is heading towards a smaller size limit asymptotically [9,10,11]. Particles in powder become unbreakable below this limit, and the energy supplied to the device is only used to crush particles whose dimensions are above it [12]. Below this limit, grinding stops. The input energy eventually dissipates as heat when the particle size distribution reaches its lower limit. Notably, the kinetic energy of the impinging ball determines the minimum dimension that can be achieved:

$$E_{ball} = \frac{1}{2} m_{ball} v_{ball}^2 \, , \qquad (1)$$

where $E_{ball}$ is the kinetic energy of a generic marble, $m_{ball}$ and $v_{ball}$ its mass and speed. The higher $E_{ball}$, the finer the ultimate particle dimension [13].

Independently of the mill's power, when crushing coarser particles, the reduction rate is nearly linear at first. As size reduction progresses, the rate asymptotically decreases and eventually drops to zero [14].

It is also important to note that the production of ceramic raw materials demonstrates a strong demand for ultrafine size, or "reactive" grades. Reactive powders readily sinter to full density at temperatures lower by 100–200°C compared to commercial grades [15,16]. Reactive powders do facilitate solid state reactions for manufacturing special products, e.g. transparent ceramic [17].

Ultrafine powders sinter essentially in the same way as coarser grain powders. However, there is a noticeable difference in the densification behavior when compared to powders that are traditionally milled. The driving force for sintering particles of any size is the reduction of the surface energy. The driving force for sintering is given by:

$$\sigma = \gamma \kappa = \gamma \left( \frac{1}{R_1} + \frac{1}{R_2} \right) \, , \qquad (2)$$

where $\gamma$ is the surface energy of the material, $\kappa$, is the surface curvature, positive for a convex surface and negative for concave, $R_1$ and $R_2$ are the principal radii of curvature. Inversely correlated with particle size, the driving power for sintering is significantly greater for ultrafine particles than for commercially available coarser powder grades [18].

The extremely low comminution process efficiency and high energy consumption have made it necessary to understand the fundamentals in order to increase efficiency and forecast the distribution of particle sizes as a function of processing variables, such as milling conditions and time.

The result of this endeavor has been a massive body of literature. While some papers have dealt with the engineering of mill equipment, the great majority of scientific effort has gone toward developing a comminution law that can connect the energy supplied to the size of the powder particles. Recently, the discrete element method (DEM) coupled with population balance formulations has greatly contributed to the advancement of understanding the ball milling process [6,19,20].

The high surface energy of ultrafine particles can be seen as a potential energy stored during powder comminution, which contributes to their high reactivity. This energy can be promptly released helping thermally activated sintering reactions lowering the energy threshold [21,22]. Following the path marked out by Stamboliadis [23,24], and by Blanc et al. [12], the present work aims at introducing the concept of energy balance between the input energy per unit of mass $E_M$ and the specific surface per unit of mass $S_M$:

$$\frac{dS_M}{dE_M} = 0 , \qquad (3)$$

and linking the achievement of a condition of balance to the minimum size that can be reached.

After a detailed description, the proposed model is validated by analyzing literature data [12] on the milling of $SiO_2$ powder, describing the case of different mill loading, and on talc powder for the case of different ball sizes. Finally, the model is exercised on experimental particle size distributions measured during the industrial preparation of a $SiO_2$ slurry in a tumbler mill.

2. Model

The fracture of a particle obviously requires energy. Higher energy absorption by the particle results in a finer average size of the particle population, according to single-particle impact fracture data. This gives rise to the concept of a particle population absorbing energy and that energy absorption persists as the size distribution moves toward ever-finer particles.

The relationship between comminution energy absorbed per unit mass and the representative size $d_r$ can be generally approximated by a differential equation [25,26]:

$$\frac{dE}{dd_r} = f(d_r) , \qquad (4)$$

where $f(d_r)$ is a decreasing function of the size $d_r$ reflecting the fact that more energy per unit mass is required as the particles get smaller.

Many formulations for the functional form of $f(d_r)$ have been offered. The earliest of these is Rittinger's law [27], which has a physical basis in the fact that the energy required for comminution is proportional to the number of chemical bonds broken in order to generate a new surface. Rittinger's law refers to the surface area and it is expressed in its integrated form as:

$$E = K_{Ritt} \cdot \left(\frac{1}{d_2} - \frac{1}{d_1}\right) , \qquad (5)$$

where $E$ is the energy imparted to the milling powder, $d_1$ and $d_2$ (with $d_2<d_1$) are the mean particle sizes at two subsequent milling steps, and $K_{Ritt}$ is a constant depending on the specific material under milling. Energy is usually expressed as $J/m^3$ or $J/kg$ depending on the units of the proportionality constant.

According to literature data, the above expression (Eq. 5) can adequately describe the particle size reduction in between the interval 20-100 μm [28,29]. At a particle size of less than 20 μm, milling studies exhibit a progressive departure from a linear energy-surface dependence. This divergence is attributed to a decrease in surface production and a decrease in grindability, which ultimately results in a higher energy demand for additional particle size reduction.

Rittinger's law posits that the energy necessary for size reduction is directly proportional to the surface area generated during the process. It further indicates that below a specific dimension (≤100 μm), the fracture energy of the powder particle, $G_p$, which is defined as the energy required to create a unit area of crack surface, remains constant and independent on the particle size: the effect of internal defects or flaws is minimal in assessing the impact stress of milling medium at the moment of fracture. Breakage is almost entirely due to body fracture, yielding approximately first-order rates [30].

The kinetic energy of the milling media appears to be inversely proportional to the limiting size that exists in fracturing, as demonstrated by experimental evidence [9]. As the size approaches the limit, the grinding efficiency asymptotically decreases from the linear Rittinger's regime, and the specific surface area eventually levels out at longer milling times.

To account for the physical particle size limit, Tanaka (quoted by Ospanov et al. [31]) and Blanc et al. [12] suggested the following modified Rittinger's crushing model:

$$\frac{dS_M}{dE_M} = \tau \cdot \left(1 - \frac{S_M}{S_{M\infty}}\right), \qquad (6)$$

where $\tau$ is the maximum comminution rate corresponding to the created surface per unit of energy, $S_M$ the specific surface area while milling, and $S_{M\infty}$ is the specific limit area at the reaching of the particle size limit. Equation (6) leads to an asymptotic behavior towards $dS_M/dE_M = 0$ and $S_M = S_{M\infty}$ after which the process practically stops.

In brittle powder milling, during the impacts on the jar walls, a ball transfers its kinetic energy to the powder as the product of loading compression ($F$) and particle elastic deformation $\Delta V_p$, being $V_p$ the volume of a generic powder particle. Up to the fracture point $F^*$, where a spherical particle usually fails by tensile fracture and releases nearly all of its stored elastic energy, the initial loading of the particle causes elastic deformation. Only a tiny amount of energy is eventually retained and converted to surface energy during crack propagation at a stress relatively close to the fracture stress [32].

The Griffith theory [33] postulates that a crack will grow when the energy released by the body per unit fracture area is greater than or equal to the increase in surface energy:

$$G \geq 2\gamma, \qquad (7)$$

where $\gamma$ (J/m$^2$) is the surface energy. In the present case $G$ can be visualized as the transferred kinetic energy at the collision site, thus Eq. (7) can be rewritten as:

$$E_M \geq \gamma \, \Delta S_M, \qquad (8)$$

where $E_M$ is the energy transferred and $\Delta S_M$ is the specific surface increase per unit of mass and already contains the factor 2 in eq. (7). When the particle size reduces, $\Delta S_M$ grows continuously, whereas $E_M$ remains constant and only depends on the milling parameters. The specific limiting area $S_\infty$ corresponds to the condition $E_M = \gamma \, \Delta S_\infty$, every impact loses its capability to crush; all the energy is dissipated into heat generated by internal friction.

In the present model the Griffith criterion is introduced, and it is applied to calculate the minimum particle size, *ergo* the maximum specific surface attainable as a function of the total impact energy $E_M$ supplied by the milling media to the powder mass unit.

In order to create an analytical kinetic formula for the mean particle size evolution, it is necessary to add a few simplifying assumptions:
a.  Energy is released to the system through a set of discrete impacts in time.
b.  The milling media are balls, all with the same size and the same speed at impacting.
c.  The contents of the mill are perfectly mixed.
d.  Particle size reduction occurs through impact breakage.

e. The material that is going to be crushed shows brittle fracturing only.
f. Particles are supposed to be sufficiently small to follow the Rittinger's regime for particle size reduction; they are also supposed to be spherical and mono-disperse.
g. The energy of the colliding media is ingested by the powder particles as elastic and plastic energy.
h. The small fraction of energy spent for crushing a particle [32] is accumulated in the powder as surface energy and the specific surface energy is independent of the particle size.
i. The powder under grinding retains a mono-disperse distribution, its evolution is being tracked through its average particle size $d_{avg}$.
j. After each collision the number of particles is expected to increase and the mean size to decrease according to the volume conservation law.
k. Energy dissipation due to particle internal friction induced by plastic and cyclic elastic deformation contributes to the poor overall process efficiency.

The quantities involved in the process are:
i. The volume of powder $V_0$ (m$^3$) trapped between the colliding bodies in an impact event; in case of dry milling, this active volume can by approximated as a cylinder, of which the area of base is Hertz's contact surface of rigid sphere in contact with an elastic half-space [34], and the height that is idealized as a bed of four layers of particles being nipped in an impact of a specified energy [35,36].
ii. The starting average particle size $d_0$ (m).
iii. The specific surface energy of the material under milling $\gamma$ (J/m$^2$).
iv. The impact energy $E_0$ (J) available at each collision.

The volume conservation in time can be written as:

$$V_0 = \frac{\pi}{6} N_0 d_0^3 = \frac{\pi}{6} N(t) d_{avg}(t)^3 \,, \qquad (9)$$

where $N_0$ is the initial number of particles of size $d_0$, and $N$ ($N \geq N_0$) the number of smaller particles with average size $d_{avg} \leq d_0$ created by grinding at a given time $t$.
According to the Rittinger's hypothesis, the energy inelastically absorbed by the powder is fully used to create new surfaces. As a function of time, the surface area per volume unit $S_V$ (m$^2$/m$^3$) can be expressed as:

$$S_V = S_{V0} + \eta \cdot \frac{E_0}{\gamma V_0} \cdot u(t) \,, \qquad (10)$$

where $u(t)$ represents the total number of collisions during milling's elapsed time.
For all kinds of milling equipment $u(t)$ can be expressed as:

$$u(t) = f \cdot t \,, \qquad (11)$$

$f$ being the mill rotation or cycling frequency, and $t$ the elapsed time.
The ratio of the energy required to generate new surfaces to the total energy input is known as the process efficiency, and the factor $\eta$ ($0 < \eta \leq 1$) represents this ratio visually. It encompasses all of the elements that release the energy input (kWh) into heat, including material internal friction, equipment mechanical friction, and interactions between powder and milling media [32].
The quantity

$$\eta E_0 / \gamma V_0 = \Delta S_V \qquad (12)$$

represents the new surface created by an impact per unit of volume, and $S_{V0}$ is the initial specific surface. Given the assumption of a mono-disperse distribution of the ground powder, the specific surface area of the particle can be represented as a function of the average particle size as follows:

$$S_V = N \cdot \frac{\pi d_{avg}^2}{V_0} = \frac{6V_0}{\pi d_{avg}^3} \cdot \frac{\pi d_{avg}^2}{V_0} = \frac{6}{d_{avg}}. \tag{13}$$

The specific energy stored as surface energy per unit volume, $E_V$ (J/m³), is a function of time:

$$E_V(t) = \gamma S_V(t) \tag{14}$$

and the pure powder's energy is:

$$E_{V0} = \gamma S_{V0} . \tag{15}$$

According to Eq. (12), the energy stored in the powder during the fragmentation process can be expressed as:

$$\Delta E_V(t) = (E_V(t) - E_{V0}) = \gamma \cdot (S_V - S_{V0}) = \eta \cdot \frac{E_0}{V_0} \cdot ft . \tag{16}$$

The incremental stored energy:

$$K_{EV} = \frac{E_0}{V_0} \cdot \eta f \tag{17}$$

is analogous to Rittinger's law in Eq. (5).
Milling energy and the overall surface vary linearly with time and the ratio $dE_V/dS_V$ is constant, in agreement with the Rittinger's model of fragmentation. It is worth noting that the proportionality constant $K_{EV}$ becomes a function of the energy transferred to the unit of powder.
When taking into account the time evolution, the energy stated in Eq. (16) experiences the same limitations as the Rittinger's law:

$$\lim_{t \to \infty} E_V = \infty, \quad \lim_{t \to \infty} S_V = \infty, \quad \lim_{t \to \infty} d_{avg} = 0 , \tag{18}$$

resulting in nonphysical outcomes.
Focusing the attention on a single fragmentation event, the volume energy transferred to powder particles of average dimension $d_{avg}$ can be expressed as:

$$E_{Vd_{avg}} = \frac{\pi}{6} \cdot d_{avg}^3 \cdot \left(\frac{E_0}{V_0}\right) , \tag{19}$$

where $E_0/V_0$ is the energy density available at the impact.
Particle fracturing generates new fresh surface $\Delta S$, and the energy associated with the newly formed surface ($E_{ns}$) can be expressed as:

$$E_{ns} = \gamma \cdot \Delta S . \tag{20}$$

The newly created specific surface, assuming that a generic mother particle of dimension $d_{avg}$ is divided into $j$ daughter particles of dimension $d_j$, can be computed as follows:

$$\Delta S = j \cdot \pi d_j^2 - \pi d_{avg}^2 . \tag{21}$$

Owing to the volume conservation, during the collision event:

$$\frac{\pi}{6} d_{avg}^3 = j \cdot \frac{\pi}{6} d_j^3 , \tag{22}$$

and the average size of the new fresh particles for a mono-disperse system is:

$$d_j = \frac{d_{avg}}{j^{1/3}} . \tag{23}$$

Substituting this value for $d_j$ in equation (22), it can be rewritten as:

$$\Delta S = \pi d_{avg}^2 \cdot (j^{1/3} - 1) = \pi d_{avg}^2 \cdot \beta_j \tag{24}$$

where $\beta_j = (j^{1/3} - 1)$.
The number of particles generated by splitting the parent powder grain is always $j \geq 2$, being $j = 2$ the minimum number of particles corresponding to a binary fragmentation. Regarding the formation of new surface, the *binary fragmentation* leads to the minimum possible surface increment $\Delta S$ during a collision event:

$$j = 2 \rightarrow \beta_2 \cong 0.26 . \tag{25}$$

The case $j$=1 means that the collision could not fracture the powder particle, consequently $\Delta S$=0.
For binary fragmentation, the Griffith's condition of Eq. (8) can be rewritten as:

$$E_{V d_{avg}} = \frac{\pi}{6} \cdot d_{avg}^3 \cdot \left(\frac{E_0}{V_0}\right) \geq \gamma \, \Delta S = \pi \beta_2 \gamma d_{avg}^2 \tag{26}$$

The case

$$E_{V d_{avg}} = \gamma \, \Delta S (j = 2) \tag{27}$$

draws great interest. It represents the energy balance in the comminution process: as a result of binary fragmentation, the instantaneous elastic energy of a powder particle equals the energy needed to generate the extra surface. Milling becomes ineffective and comminution stops progressing.
The ultimate particle dimension $d_{lim}$ achievable by milling can be estimated from equation (27) as:

$$E_{V d_{lim}} = \frac{\pi}{6} \cdot d_{lim}^3 \cdot \left(\frac{E_0}{V_0}\right) = \pi \beta_2 \gamma d_{lim}^2 . \tag{28}$$

Thence:

$$d_{lim} = 6\beta_2 \cdot \gamma \cdot \frac{V_0}{E_0} \tag{29}$$

where the factor $6\beta_2 \cong 1.56$.
Physical characteristics, particularly the specific surface energy of the milled powder and the energy density of the milling media related to their kinetic energy at impact, dictate the final particle size reduction.
Notice that neither the milling time nor the process efficiency ($\eta$) concur to the smallest achievable particle size. The process efficiency only affects the time required to reach the final equilibrium size.

The formulation of a kinetic equation describing the evolution in time of the grinding process needs to fulfill the following constraints:
a) The equation complies with Rittinger's model (Eq. 15) for the first stage of comminution: the average reduction in particle size is directly proportional to the milling time (constant grinding rate).

b) At a given time, the ultimate particle size and energy equilibrium are asymptotically reached by the reduction rate gradually decreasing during the comminution process. Then, because of internal and external friction mechanisms, all of the mechanical energy used in the milling process is lost as heat.

Following the approach proposed by Blanc et al. [12], the nonlinear evolution of the specific surface can be justified through the introduction of an additional damping factor accounting for the energy budget needed to form a new fracture surface.

The number of particles, their dimension and the powder specific surface are intimately interrelated. For the present purpose it is more convenient to focus on the specific surface, since it is an experimental observable that is easily measured, and it is insensitive to the particle morphology.

To examine the specific surface evolution as a function of the milling time, Eq. (10) can be differentiated to define the comminution kinetics. Recalling that $u(t)=f \cdot t$:

$$dS_V = \eta \cdot \frac{E_0}{\gamma V_0} \cdot f dt \tag{30}$$

or rearranging:

$$\frac{1}{\eta f} \cdot \frac{dS_V}{dt} = \frac{E_0}{\gamma V_0} . \tag{31}$$

An attenuation factor can be added to Eq. (31) to allow for the continual decrease of the grinding rate, or, consequently, the rate of generation of new surface area, as well as the achievement of an upper limit for the final specific surface:

$$\frac{1}{\eta f} \cdot \frac{dS_V}{dt} = \frac{E_0}{\gamma V_0} \cdot \left(1 - \frac{S_V}{S_{Vlim}}\right) \tag{32}$$

where the limiting specific surface per volume unit, $S_{Vlim}$, is:

$$S_{Vlim} = \frac{6}{d_{lim}} = \frac{E_0}{\beta_2 \gamma V_0} . \tag{33}$$

Integrating Eq. (32) between $S_{V0}$, specific surface of the initial powder and $S_V$, the specific surface at a generic time, and between $t=0$ and a generic time $t$ and rearranging the terms of the equation, one obtains:

$$S_V(t) = S_{Vlim} - (S_{Vlim} - S_{V0}) \cdot exp\{-\frac{E_0}{\gamma V_0 S_{Vlim}} \cdot \eta f t\} . \tag{34}$$

At the comminution start, $t=0$ and $S_V = S_{V0}$, while for any milling time $S_V > S_{V0}$. Extending the milling time, $t \gg 1$, $S_V \approx S_{Vlim}$. Recalling Eqs. (29) and (33), Eq. (34) can be alternatively formulated as:

$$S_V(t) = S_{Vlim} - (S_{Vlim} - S_{V0}) \cdot exp\{-\beta_2 \cdot \eta f t\} . \tag{35}$$

The first derivative with respect to time of Eq. (35) represents the rate of creation of new specific surface:

$$\frac{dS_V(t)}{dt} = (S_{Vlim} - S_{V0}) \cdot \beta_2 \eta f \cdot exp\{-\beta_2 \cdot \eta f t\} . \tag{36}$$

At the very beginning of the milling process, $t \to 0$:

$$\lim_{t \to 0} \frac{dS_V(t)}{dt} = (S_{Vlim} - S_{V0}) \cdot (\beta_2 \eta f) . \tag{37}$$

Taking into account that $S_{V0} \ll S_{Vlim}$ and considering Eq. (29), the slope of $S_V(t)$ at $t=0$ is:

$$K_{SV}' = \left.\frac{dS_V}{dt}\right|_{t=0} \cong S_{Vlim} \cdot (\beta_2 \eta f) = \frac{E_0}{\gamma V_0} \cdot \eta f \,. \tag{38}$$

Conversely, for an extended milling time $t>>1$:

$$K_{SV}' = \left.\frac{dS_V}{dt}\right|_{t\gg 1} \cong 0 \,. \tag{39}$$

Equation (35) fulfills the expressed requirements:
At the beginning of the process it reduces to Rittinger's equation and the comminution rate $K_{SV}'$ becomes identical to the Rittinger's constant in Eq. (10) and (17). The comminution rate gradually falls as the milling time increases, eventually asymptotically tending towards zero, which is the steady state equilibrium condition for longer milling times.
Assuming $S_{Vlim} \gg S_{V0}$, Eq. (35) can be rewritten in a dimensionless form as:

$$x(t) = \frac{S_V(t)}{S_{Vlim}} = 1 - \frac{(S_{Vlim}-S_{V0})}{S_{Vlim}} \cdot exp\{-\beta_2 \cdot \eta ft\} \cong 1 - exp\{-\beta_2 \cdot \eta ft\} \,. \tag{40}$$

Though the processes are very different, it is worth noting that Eq. (40) mathematically resembles the Johnson-Mehl-Avrami-Kolmogorov (JMAK) equation defining the kinetics of phase changes in condensed systems [37]:

$$y(t) = 1 - exp\{-bt^n\} \,, \tag{41}$$

where $y(t)$ is the fraction of material that is transformed as a function of time, and $b$ and $n$ are constants. In the present case $n=1$.
With the aid of the set theory formalism, certain similarities and parallels between the phase transformation and the comminution progress may be inferred.
Actually, the ensemble of the powder particles contained in the volume $V_0$ can be interpreted as compact set $P$ composed of two subsets: $A$ containing the particles having a specific surface $S_A < S_{lim}$, and $B$ having a specific surface $S_B \geq S_{lim}$, so that $A \cup B = P$, and $V_A + V_B = V_0$, $V_A$ and $V_B$ being the volume of the subsets $A$ and $B$, respectively.
Under the impact compression, the particles $p_i \in A$ are forced into a metastable state. They react by splitting in two or more particles $p_i \to p_j$ with $j = 1, \ldots n$. Any resulting particle that has a particular surface area larger than or equal to $S_{lim}$ is placed in subset B, $p_j \in B$, which is in equilibrium under compression. The parent particle $p_i$ vanishes, and the kinetics of depletion of the subset $A$ and population of the subset $B$ continue as a phase transformation controlled by Eq. (41). Ultimately, $A$ becomes an empty set ($A=\emptyset$), $B$ an improper subset of $P$, ($B \subseteq P$), and comminution stops.

3. Test of the model
Obviously, a model asks for validation against experimental results. Here three cases have been examined, the first two from literature data [12,11], the last one from the industrial production of ultrafine silica slurry.
The first two examples have been selected because they focus on a peculiar feature of the milling process. The first example, a guideline for the present research [12], deals with the dry milling of silica powder. In this experiment an oscillating ball mill was used. The main feature of the milling equipment is the use of a single milling sphere, and the milling outcomes are investigated depending on oscillation frequency and starting quantity of powder.
The second test refers to the application of a high energy planetary mill on talc powder [11]. The study examines the effect of varying the milling ball size, all the other parameters kept constant. Accordingly, the ultimate specific surface should change, the amount of energy delivered to the powder being proportional to the ball mass. The last case refers to the optimization of the wet milling process in a tumbler mill aimed at increasing the green density of silica cast manufactures.

*3.1. Case 1: Dry grinding of silica by a laboratory mill*

Blanc et al. [12] performed an experiment on the dry milling of quartz powder utilizing a Retsch MM 400 vibro-mill equipped with 50 mL milling jars. Each jar is initially filled with a mass $m_{SiO2}$ of raw material and a single ball. The motion of the jar follows a quasi-straight line with a periodic signal:

$$x_{jar} = A \cdot sin(2\pi f t) , \qquad (42)$$

where *f* is the oscillation frequency and *t* is time. The stroke amplitude of the jar is *A* = 7.20 mm independently of frequency.

The maximum velocity of the jar that coincides with the maximum velocity of the ball, is:

$$v_{jar}^{max} = 2A\pi f = v_{ball}^{max} \qquad (43)$$

The maximum impact energy is:

$$E_{impact} = \frac{1}{2} m_{ball}\left(v_{ball}^{max} + v_{jar}^{max}\right)^2 = 8 m_{ball}(A\pi f)^2 \qquad (44)$$

where $m_{ball}$ is the mass of the milling sphere, a 25 mm diameter steel ball in the present case.

Here, the set of experiments at a fixed frequency, 20 Hz with increasing values of the powder mass (6, 13, 17 and 20 g) are considered.

The model predicts that all trials conducted at the same frequency (20 Hz) will produce a powder with the same final specific surface and, as a result, the same limit particle size. Throughout all of the trials, the impact energy $E_0 = E_{impact}$, the impacted volume $V_0$, and the milling ball remain unchanged:

$$S_{Vlim} = \frac{6}{d_{lim}} = \frac{E_{impact}}{\beta_2 \gamma V_0} . \qquad (45)$$

The mass of powder processed at each collision is:

$$m_{impact} = \rho V_0 \qquad (46)$$

where, $V_0$ is the Hertzian impact volume and $\rho$ the density of the milled powder.

The milling time required for the whole powder batch to attain the ultimate specific surface, namely the steady-state condition, can be estimated as:

$$t_{tot} = \frac{m_{SiO2}}{m_{impact}} \cdot \frac{1}{f} \qquad (47)$$

In the following analysis, the last sampling time in the milling runs has been taken as representative of the total time.

The mass of powder trapped between the colliding ball and the jar wall, $m_{impact}$, estimated according to Eq. (46), is shown in the fourth row of Table 1. Equation (45) predicts that the final value of the specific surface should tend toward the same value for all the mass values under investigation since the impinging ball's velocity is constant. The ball radius, ball velocity at collision, and material deformation all affect the volume of impacted powder at each stroke [34]. The impact volume $V_0$ ought to be independent of the mass loading because all of the parameters are the same for every run. Consequently, the mass of powder processed at each collision should be constant, and the mass-to-collision ratio, the quantity of powder trapped at each collision, may be inferred directly from the trials and is a function of the time $t_{tot}$ required to reach the asymptotic milling regime.

Table 1. Calculated mass of powder trapped between the colliding ball and the jar wall, energy density and process efficiency optimized by non-linear least squares fitting (see Fig. 1) and additional calculated quantities. Stroke frequency is 20 Hz for all runs.

| Mass of silica powder $m_{SiO2}$ (g) | 6 | 13 | 17 | 20 |
|---|---|---|---|---|
| Maximum milling time in the experiments $t_{tot}$ (min) | 29 | 65 | 89 | 111 |
| Number of collisions | 34800 | 78000 | 106800 | 133200 |
| Mass-to-collisions ratio $m_{impact}$ (g × 10$^4$) | 1.72 | 1.67 | 1.59 | 1.50 |
| Optimized energy density $[E_0/V_0]$ (kJ/m$^3$) | 231 | 219 | 232 | 238 |
| Optimized efficiency $\eta$ | 7·10$^{-4}$ | 4·10$^{-4}$ | 2·10$^{-4}$ | 2·10$^{-4}$ |
| Calculated limiting size $d_{lim}$ (μm) | 2.7 | 2.9 | 2.7 | 2.6 |
| Calculated slope of $S_M$ versus time, $K_{SM}$ (m²/kg/min) | 184 | 104 | 73 | 50 |
| $\eta \cdot m_{SiO2} \cdot 10^4$ (g) | 40 | 55 | 50 | 40 |
| $S_{M\infty} = 6/(\rho \, d_{lim})$ (m²/kg), Eq. (45) | 837 | 794 | 843 | 865 |
| $S_{M\infty}$ experimental (m²/kg) | 800 | 798 | 885 | 900 |
| Correlation coefficient $r$ | 0.99 | 0.98 | 0.99 | 0.99 |

Comparing the calculated ratios for the different runs, a slight decrease of the mass-to-collision ratio is observed when increasing the powder content. As known, high powder content may generate friction inside the jar, thus slightly reducing the milling efficiency and the quantity of processed powder per unit of time [38]. The kinetics of the grinding process has been analyzed by least squares fitting Eq. (34) to the experimental data measured on the milling experiment performed at 20 Hz with increasing powder mass. The results are shown in Figure 1 and Table 1.

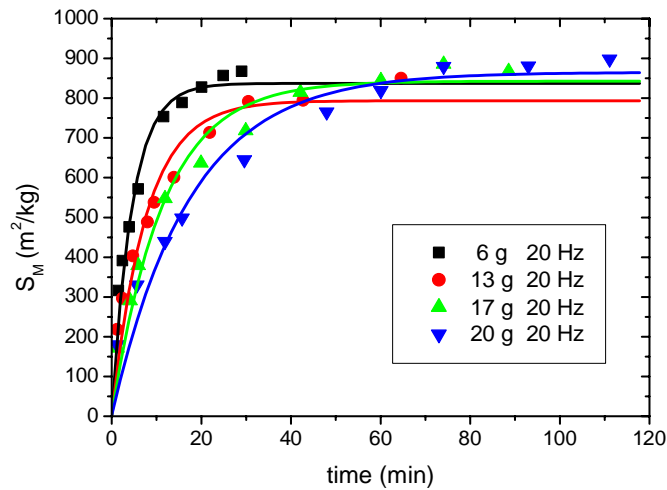

Fig. 1. Non-linear fit by eq. (35) of the experimental milling data by Blanc et al. [12] at a frequency of 20 Hz. The specific surface per unit mass is measured by laser diffraction (Mie's scattering).

The curve fitting has been implemented taking the energy density and the process efficiency as free parameters, the specific surface energy has been set at $\gamma \approx 0.40$ J/m$^2$ according to literature data [39]. The experimental data are the specific surface area as a function of the milling time and measured by laser diffraction (LD) technique (Mie's scattering).

As expected, in accordance with the proposed model, the energy density is almost constant and independent on the quantity of the powder charge. The returned values of the energy density $E_0/V_0$ are in some extent lower than the value calculated from Eq. (44) and Eq. (46): $E_0/V_0 \approx 280$ kJ/m$^3$.

The $E_0/V_0$ best fit findings are convincingly matching the values estimated from the milling parameters, taking into account all potential sources of error, particularly the accurate estimation of the relative speed of the jar and milling ball at the impact instant.

The best fit procedure provides process efficiency values in the range $\eta \approx 2 \cdot 10^{-4} \div 7 \cdot 10^{-4}$. Observing that the product between the efficiency and the mass of powder ($\eta \cdot m_{SiO2}$) is nearly constant leads to conclusions about the constancy of the mass trapped between the colliding ball and the jar wall that are similar to those reported for Tab. 1. According to Eq. (35), efficiency acts only on the characteristic time of the process. In fact, the time needed to achieve a fraction $x$ of the asymptotic specific surface is:

$$t_x = \frac{-\ln(1-x)}{\beta_2 \cdot \eta f} \ . \tag{48}$$

Besides mill mechanical friction and ball-powder rubbing friction, it is important to recall that only a very small fraction of the elastic energy transferred to the powder at each impact is consumed to form new fracture surfaces [32]. An estimate of the highest conversion of the kinetic energy into fracture surface energy can be evaluated from the slope of $S_M(t)$ at the origin, where comminution obeys the Rittinger's law [Eq. (38)], and the kinetic energy imparted to the powder in the unit of time:

$$\eta_{max} = \frac{\left.\frac{dS_V}{dt}\right|_{t=0} \cdot \gamma \cdot m_{SiO2}}{[8\, m_{ball}\, (60 \cdot A\pi f)^2]} . \tag{49}$$

where the factor 60 at the denominator accounts for the time unit (1 min) used in the slope calculation. Inserting the values of both experimental and calculated quantities: $\eta_{max} \approx 1.1 \cdot 10^{-3} \div 1.4 \cdot 10^{-3}$. The quantity $\eta_{max}$ represents the maximum percentage of elastic energy that can be converted into fracture surface energy in absence on any friction.

As already mentioned, $d S_M/dt$ is monotonically decreasing down to a steady state $d S_M/dt \cong 0$ for long milling time. The highest efficiency is always reached at the beginning of the comminution process.

Blanc et al. also analyzed the evolution of the particle size distribution depending on the milling time. In Figure 2 the particle size distribution (PSD) relative to the sample 13 g and milled for 65 minutes at 20 Hz captured from the original figure in Blanc's paper is shown. The decomposition of the particle size distribution profile retrieved from reference [12] into the sum of two log-normal components allows to estimate the smallest average dimension of the SiO$_2$ particle (red line) at $d_{lim} \approx 2.7$ μm, in perfect agreement with the ultimate specific surface value.

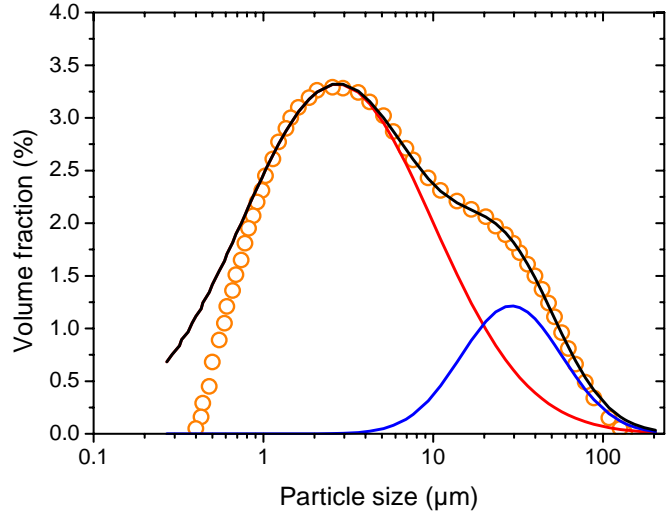

Fig. 2. PSDs measured by laser diffraction on the 13 g sample milled at a frequency of 20 Hz by Blanc et al. [12] (open symbols). The experimental curve has been decomposed into the sum of two log-normal distributions (red and blue lines) to separate the components.

*3.2. Case 2: Wet milling of talc powder using a high energy ball mill*
Kim et al. [11] investigated the facility of producing high quality and ultrafine talc powder by high energy ball milling. Ultrafine talc powder is a valuable additive in several industrial fields (paint, paper, rubber, ceramic, and polymeric manufacturing) and in the chemical and pharmaceutical industry.
Grinding talc platelet-like powder is not straightforward. Delamination and exfoliation of the layered structure in the direction of the (001) crystal surface are the predominant mechanisms of comminution, thus multiplying the number of platelets with reduced thickness and comparable surface.
Aimed at attaining ultrafine powder, they used a high energy ball mill (EMAX, Retsch-Germany) working on the same principle of a planetary ball mill and able to work at very high rotation speed (for more experimental details please refer to the original paper). In their study they analyzed the influence of the milling media size on the final outcome.
The EMAX mill belongs to the planetary mills family and the jar has a very large volume (125 mL) compared with device in case 1. Planetary mills slingshot the milling media against the jar wall with a velocity which is almost independent of their mass and controlled by the rotation speed of the mill platform [40,41]. Due to the jar dimension, the impact can be well approximated as the contact of an elastic sphere with an elastic half-space.
Using $ZrO_2$ balls of three increasing diameters—0.1, 1, and 2 mm—they mill talc powder in their experiment while maintaining consistent conditions and settings. Specifically, a volume ratio of 2:3:4 has been implemented for zirconia balls, water, and talc powder. The verification of the final powder specific surface's dependence on the energy density $E_0/V_0$ is a relevant use of this experiment.
Let $E_0$ (J) be the kinetic energy of the colliding ball, since $v_{ball}$ is approximately constant and the milling media are of the same material, the impact energy variation is only determined by the mass $m_{ball}$ of the used milling media:

$$E_0^{r=0.1} : E_0^{r=1} : E_0^{r=2} = m_{ball}^{r=0.1} : m_{ball}^{r=1} : m_{ball}^{r=2}. \tag{50}$$

The volume of the nipped powder or active volume $V_0$ (m³) is defined by the Hertzian surface of contact and a layer of powder $h$:

$$V_0 = \pi a^2 h \tag{51}$$

where *a* is the is the radius of the contact area:

$$a = \left(\frac{3\,F\,r_{ball}}{4\,Y_m}\right)^{1/3}, \tag{52}$$

where *F* is the applied load, $r_{ball}$ the ball radius and the mean elastic modulus $Y_m$ is:

$$\frac{1}{Y_m} = \frac{1-\nu_1^2}{Y_1} + \frac{1-\nu_2^2}{Y_2}, \tag{53}$$

$Y_1$ and $Y_2$ being the elastic moduli and $\nu_1$ and $\nu_2$ the corresponding Poisson's ratios of the milling balls and jar walls. The applied load can be expressed as $F = m_{ball} \cdot v_{ball} \cdot \Delta t$. Being $v_{ball}$ and subsequently $\Delta t$, the impact duration, steady, the applied load ratio depends only on the mass of the milling media.
Finally, settling for determining the sole ratios of the ultimate specific surface per mass unit as a function of ball size, $S_{M\,lim}^{r=2} : S_{M\,lim}^{r=1} : S_{M\,lim}^{r=0.1}$, recalling that $S_M \propto E_0/V_0$, all the material constants and $\Delta t$ are canceling each other out. Calculation is substantially simplified and the surfaces per mass unit are proportional to $m_{ball}^{1/3} \cdot r_{ball}^{-2/3}$.
By considering the proportionality between mass and cubed radius of the balls through the density of the grinding medium, the final relationship is:

$$S_M \propto r_{ball}^{1/3}. \tag{54}$$

The experimental results reported by Kim et al. [11] are summarized in Figure 3 and Table 2.

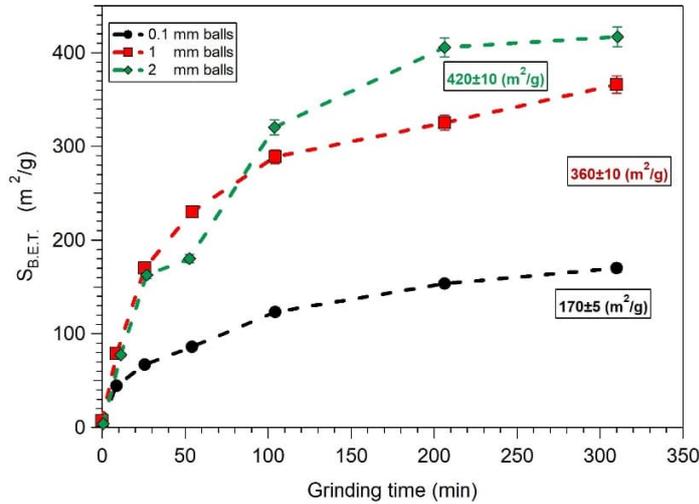

Fig. 3. Specific surface area as a function of the milling time for three different $ZrO_2$ ball sizes. The mill rotation speed is 2000 rpm.

Table 2. Main results on talc milling with a high energy ball mill. All data are taken from the original paper of Kim et al. [11]. The last two columns show the ratio of the BET specific surface area measurements. The second to last column is the ratio among experimental values, the last one the calculated rate according to the present model. Note: the equivalent sphere diameter is $ESD = 6/(\rho\, S_M)$, where $\rho = 2800$ kg/m$^3$.

| Ball Size (mm) | $S_{BET}$ (m$^2$/g) Raw powder | $S_{BET}$ (m$^2$/g) After 360 min. | Equivalent Sphere Diameter (nm) | $S_{BET\,lim}^{r=2} : S_{BET\,lim}^{r=1} : S_{BET\,lim}^{r=0.1}$ experimental | $S_{BET\,lim}^{r=2} : S_{BET\,lim}^{r=1} : S_{BET\,lim}^{r=0.1}$ calculated |
|---|---|---|---|---|---|
| 2 | 6.1 | 420 ± 10 | 5 | 2.4 ± 0.2 | 2.7 |
| 1 | 6.1 | 360 ± 10 | 6 | 2.1 ± 0.2 | 2.2 |
| 0.1 | 6.1 | 170 ± 5 | 12 | 1.0 | 1.0 |

The agreement between the experimental results and the values determined by an independent procedure implies that the $E_0/V_0$ ratio in Eq. (33), in turn, controls the maximum specific surface area that can be achieved or, consequently, the smallest average particle size.

The experimental lower ratio for the 2 mm ball size indicates a slightly reduced grinding efficiency. The reduced number of milling balls is probably the cause of this variation. The smaller number of contact points of the larger balls (a negative source) partially frustrates the enhanced kinetic energy of the colliding media, which is a positive source for milling efficiency. Also the higher scatter in the specific surface area curve for the largest size (2 mm) hints that the ball size and number of is not the most favorable compromise moving away from the ideal conditions [42].

The elastic energy density that the powder enveloped in the active volume $V_0$ during the impact can be expressed as the ratio $E_0/V_0$ [J/m$^3$]. Therefore, $E_0/V_0$ turns into a physical quantity that is dependent on the material's characteristics and the milling settings, enabling the prediction of the milling action's ultimate limit from an energy perspective.

*3.3. Case 3: Industrial wet milling of silica (silica slurry)*

The last case dealt with is the production of ultrafine silica-based slurry (82.5% solid component, 17.5% water) for slip casting. The slurry production process begins with fragmenting and crushing high purity quartz cullet, followed by long-time wet milling, up to 48 hours, in a drum ball mill. Since the industrial application of the silica slurry, the progress of the batch cycle milling is monitored and completed once a set average particle size measured by laser diffraction is reached.

The milling equipment consists of a drum ball mill with an inner diameter of 0.6 m rotating at a frequency of 65% the critical speed. The milling medium is composed of 1-inch Al$_2$O$_3$ balls with a powder-to-balls mass ratio equal to 0.32. Deionized water has been added to attain a slip with a proper density $\rho \approx 1.8$ g/cm$^3$.

Among the several runs needed to optimize the slip preparation, the two investigated runs tagged 107P and 206P refer to the concluding optimization phase. Although milling did not reach the ultimate steady state particle size or, in turn, the final specific surface, nonetheless the powder sampling in time, is appropriate for validating the proposed model.

Particle size distribution curves, volume fraction vs. particle diameter (spherical approximation), are shown in Figure 4. The specific surface of the powder is calculated from the PSDs assuming spherical silica particles ($\rho = 2650$ kg/m$^3$). The first sampling was made after 20.75 (107P) and 24 (206P) hours of milling. At this time, a conspicuous amount of powder (volume fraction) has already reached the lowest allowable dimension. In both milling runs, the right tail of the curves shows the progressive disappearance of the largest particles. The peak relative to the smallest particle size is located at about 5 μm.

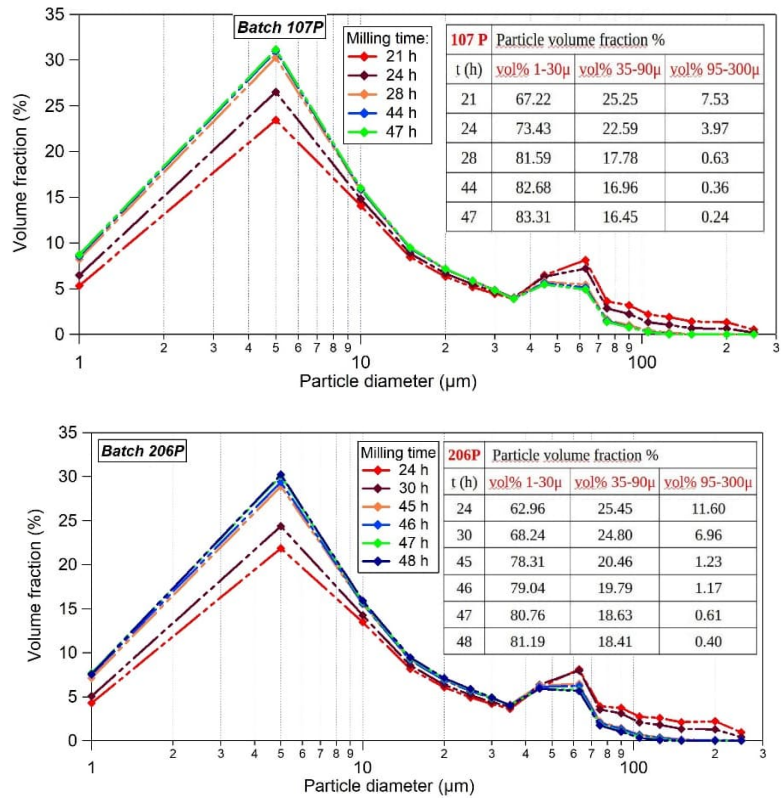

Fig. 4. Particle size distribution (volume %) as a function of the milling time. Top: batch 107P; Bottom: batch 206P. Both batches have the same starting composition and have been milled under the same nominal conditions.

When comparing the PSDs, it seems that the coarse-grained particles are being broken up into smaller pieces by milling first. The milling action moves to the next finer particle class and repeats the procedure after the coarse particles are completely destroyed. Simultaneously, crushed particles that reach the limiting size harden against breaking and dissipate mechanical energy into thermal energy.

In their work on the relationships between particle size and fracture energy, Yashima et al. [43] investigated the relationships between fracture energy and compressive strength of sphere on the basis of fracture mechanics applied to the crushing of single spheres. They showed that fracture energy and compressive strength of a sphere made of brittle material increases as the particle size decreases. Similarly Tavares et al. [44] proved that a decrease in the particle size resulted in a shift of the distributions of particle fracture energy of quartz to higher values.

The impact energy $E_0$ in a tumbler mill is dependent on several factors, such as the size of the drum, the mass of the milling medium, the design of the lifter, the cascading or cataracting regimes, the filling ratio, and so on. The discrete element approach simulation of the charge motion provides more precise information on the impact energy of the milling media [45].

Figure 5 shows the number of impacts as a function of the mean energy of a single colliding ball. The histogram, originally calculated with the DEM method by Datta et al. [45], has been adapted to the present experimental milling conditions. Crushing coarse particles necessitates a lower impact energy of the colliding balls and, according to DEM modeling, the number of collisions in unit of time is inversely proportional to the impact energy of the milling media. During milling, coarse powder particles are hammered hard and break into smaller bits, which causes them to vanish first (see Fig. 4). Particle fracture strength increases with increasing size reduction, but the fraction of impacting balls with the required energy decreases by almost two orders of magnitude.

Stirring and attrition mills were developed on an industrial scale as a result of the tumbler mill's yield drastically decreasing towards lower particle sizes by design.

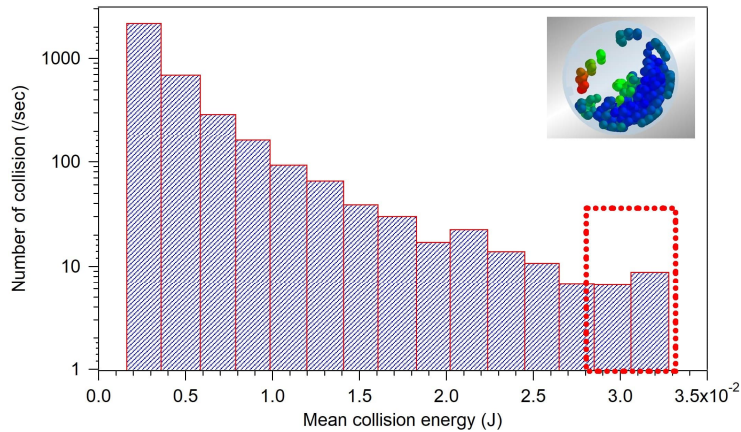

Fig. 5. Tumbler mill DEM simulation [45]. Number of collisions versus mean collisions energy. The dashed red rectangle points to the highest collision energy leading to the finest achievable particle dimension. The frequency of collision drops of two orders of magnitude towards the high energy collision side.

As mentioned above, the energy density $E_0/V_0$ is the principal quantity that determines the limit of comminution. The situation becomes more complicated in the case of wet milling. The slurry collects in the bottom of the drum, the impinging balls are splashing down in the slurry, the estimate of the impact surface (ball against drum wall) and of the active volume remains, in this case, uncertain. For these reasons Eq. (33) has been fitted to the two sets of experimental data (see Figure 6) letting the energy density $E_0/V_0$ and the efficiency $\eta$ as free parameters to be optimized by the least squares method [46].

Given that the ball milling feed in both tests is silica powder, the obtained values (see Table 3) have been closely examined and contrasted with case 1. Equation (33) well describes the accretion of the specific surface as a function of the milling time as confirmed by the high correlation coefficient. The calculated minimum particle size (spherical approximation) is really close to the value measured on the particle size distribution curve, which is of course lower than the average size measured by laser diffraction and due to the long tail towards the larger size of the particle distribution.

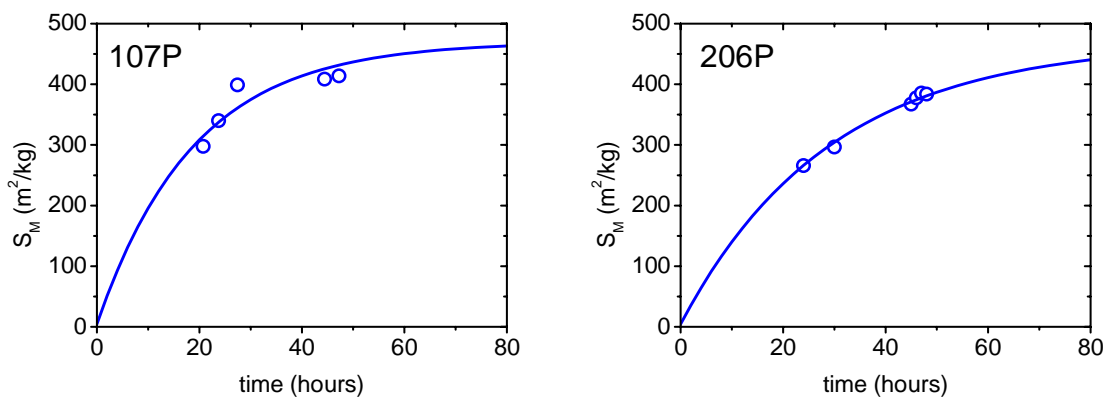

Fig. 6. Least squares fit to the experimental data sets. Left: batch 107P, right: batch 206P. The fitting results are shown in Table 3.

Table 3. Values of energy density $E_0/V_0$ and efficiency $\eta$ optimized by non-linear fitting. The minimum particle size and the maximum specific surface are calculated from $E_0/V_0$ via Eq. (33). The last column represents the average particle size measured by Laser Diffraction (LD) and used as an industrial parameter for assessing the readiness of slurry.

| Batch | $d_{lim}$ (μm) | $S_{M\,lim}$ (m²/kg) | $K_{SM}$ (m²/kg/h) | Efficiency $\eta$ | $E_0/V_0$ (kJ/m³) | Correlation Coefficient | $d_{average}$ (LD) (μm) |
|---|---|---|---|---|---|---|---|
| 107P | 5.1 | 440 | 27.3 | 9·10⁻⁵ | 122 | 0.903 | 7.7 |
| 206P | 4.6 | 495 | 15.2 | 4·10⁻⁵ | 136 | 0.996 | 8.4 |

It is worth comparing the values in Table 3 with the results previously obtained by the mono-sphere milling equipment (see Table 1). To improve reading and facilitate the comparison, Table 4 has been assembled.
It shows that:
- The $E_0/V_0$ ratio, the energy density of the powder particles gained during a collision event, scales perfectly with the smallest dimension reached by the particles and, in turn, with the largest attainable specific surface.
- The efficiency of the drum mill used for silica slurry pilot production is curtailed by one order of magnitude with respect to the single sphere oscillating mill. Actually, the single sphere vibro-mill used by Blanc et al. is always delivering the same amount of energy per collision, being the number of collision constant in time. In other words the efficiency remains stable during milling. Conversely, the efficiency of a drum mill is high with coarse grained powder, but it progressively dwindles and drops by two orders of magnitude as the reduced size of the powder asks for higher collision energy (see Figure 5).

It is important to note the actual difference between dry milling in a tiny laboratory vibro-mill and wet milling in a pilot scale tumbler mill. The largest specific surface, or the ultimate particle reduction, has a substantial correlation with the $E_0/V_0$ ratio; this correlation shows that these quantities are physically linked and independent on milling mechanics.

Table 4. Comparison of the result from the analysis of Blanc et al.'s data and the outcomes from the wet milling of silica slurry. In curly brackets the ratio of the characteristic quantities: $E_0/V_0$, $d_{lim}$ and $S_{M\,lim}$, Blanc's data being the reference.

| Sample | $E_0/V_0$ (kJ/m³) | $d_{lim}$ (μm) | $S_{M\,lim}$ (m²/kg) | Efficiency $\eta$ |
|---|---|---|---|---|
| Samples 6-20 g at 20 Hz [12] | ≈ 230  {1} | ≈ 2.7  {1} | 840  {1} | ≈ 5·10⁻⁴ |
| 107P | 120  {0.53} | 5.1  {0.53} | 441.7  {0.53} | 9·10⁻⁵ |
| 206P | 135  {0.59} | 4.6  {0.59} | 494.5  {0.59} | 4·10⁻⁵ |

4. Discussion
Throughout the literature there are many discussions about a size limit in comminution processes. Knieke's Ph.D. thesis gives a really comprehensive review of the matter [47].
Formally, the model provided here is the same as those previously demonstrated by Tanaka [31] and Blanc [12]; for details, see Eq. (6). The key innovation is the interpretation of the impact damping factor $(1 - S_M/S_{M\infty})$. The current method interprets the limit of the comminution process as the ground powder's energy equilibrium with the transiently higher energy state created by the impact of the milling media. Comminution is understood as a chemical reaction that results from collisions and creates new surface through the breakdown of atomic bonds. When the surface energy of a powder particle ($\gamma \cdot S_p$) equals the elastic energy ($E_p$) of the particle under impact, equilibrium is established.

According to Stamboliadis [24], the specific surface energy of the powder can be thought of as a potential energy that the system has stored. As a result, the comminution process stops working as soon as the powder potential energy equals or exceeds the elastic energy connected to an impact.

The difference between the powder potential energy and the impact elastic energy becomes the driving force of the process. The damping effect now describes the gradual and monotonic efficiency decrease to the final stalling.

The experimental PSDs of Figs. 2 and 4 show that, at the microscopic level, the modified Rittinger equation deviates from linearity as soon as a sizable volume proportion of particles already at the equilibrium size is generated.

The growth rate of the specific surface as a function of milling time has the same mathematical form as the JMAK equation that describes the kinetics of phase transformations in condensed systems, introducing the idea that comminution is an activated reaction (see Eqs. (40) and (41)). At the same time, a unitary exponent ensures the absence of the inflection point, in agreement with Rittinger's law.

It is important to underline that only brittle materials showing fragile fracturing behavior have been analyzed so far. The extension to ductile material still needs to be assessed, taking into account the energy associated with the generation of dislocations and stacking faults.

The size limit in comminution processes has been the subject of a great deal of research. For example, see Blanc's work [12], the mentor paper of the current study, and the references therein. The grinding limit has generally been attributed to limitations in the mechanics of the process, such as the powder coating of the grinding media, the cushioning effect caused by air trapping [48], the viscosity during wet milling [49], and limitations in the material, such as agglomeration [50], increased fracture strength with particle size reduction [51], and the existence of a threshold dimension below which further breakage is not possible [52,53] or material plastic deformation, likely due to local fusion at the impact [54].

Polymorphic transformations [55] and pseudo-amorphization [56] resulting from structural instability may facilitate energy dissipation beyond the comminution threshold.

The impact energy density $E_0/V_0$, related to the milling conditions, and the specific surface energy, related to the material properties, are the identifying factors in the energy balance approach that is being proposed here. The grinding limit, defined as the smallest achievable particle dimension $d_{lim}$ or, equivalently the largest specific surface per volume unit $S_{V\,lim}$, is attained when specific surface energy is in equilibrium with the temporary energy density within a collision event:

$$S_{Vlim} \cdot \beta_2\, \gamma \leftrightarrow E_0/V_0 \tag{53}$$

In each experiment the quantity $S_{Vlim} \cdot \beta_2\, \gamma$ is an extensive variable, while $E_0/V_0$ is an intensive variable. Regardless of the route taken to reach the ultimate state, the equilibrium condition determines the ultimate particle size or the large specific surface.

The data gathered from powder processing under radically different experimental setups — dry vibro-milling in the first instance, and wet drum milling in the second — shows that milling equipment and conditions have no incidence on the end product (see Table 4). The ultimate fineness is solely determined by the value of $E_0/V_0$. Talc powder ground in a high energy ball mill further demonstrates that Eq. (55) still holds true for materials that are exfoliated rather than crushed to a nanoscale size, retaining their crystalline structure [57].

The energy balance approach takes into account the increasing fracture energy with decreasing particle size since it suggests that a larger energy density is needed to initiate particle cracking.

Particles that agglomerate during milling and change and enlarge their observable size, may give rise to some controversy. The final equilibrium state could be puzzled by the presence of large agglomerates.

When the ultrafine particles agglomerate during grinding, there are noticeable differences between the physical absorption surface area analysis and the particle size evaluation using laser diffraction or scanning electron microscopy. Optical methods are sensitive to the whole size of the agglomerates. On the other hand, weak interactions between the constituent particles of the agglomerate prevent strong necks from forming, allowing nitrogen gas to enter the agglomerate and be absorbed on the surface of the particles. This is why BET adsorption analysis is less impacted by agglomeration than the optical method. Each particle retains its original surface, consequently $S_{agg} = \sum_i S_{P_i}$, where $P_i$=1,..., $n$ are the particles forming the agglomerate. The quantity $S_{Vlim} \cdot \beta_2 \gamma$, the total surface energy remains almost uninfluenced by the presence of agglomeration [11,12,58].

The major restriction of the proposed model is its applicability only to brittle materials showing fragile fracture. The energy balancing approach begins with the supposition that the powder being ground is gradually storing energy as specific surface energy, all the way up to offsetting the impact energy.

For instance, when using high energy ball milling for mechanical alloying (MA), the balance established between the fracturing and cold-welding events during MA determines the steady state condition and the minimum grain size of the powder. The value of $d_{lim}$ achievable by milling is regulated by the competition between the plastic deformation via dislocation motion that tends to decrease the grain size, and the recovery and recrystallization behavior of the material that tends to increase the grain size. Mechanical alloying facilitates various phase transformations, including the promotion of solid solutions, intermetallic reactions, structural destabilization, and the formation of amorphous phases. These transformations are accompanied by significant energy changes, either through absorption or release, which can influence the comminution path. Thence, the proposed energy balance method model becomes inadequate to describe this branch of powder ball milling [2,59].

5. Conclusions

The current work proposes a novel formulation for characterizing the last phase of comminution in ball milling. The approach that is presented is a reinterpretation of the Rittinger's equation with an attenuation factor added to account for the grinding rate gradually slowing down, or creating new surface, until the final leveling off that indicates the final grinding limit. The concept of attenuation factor refers to the point at which the specific surface energy obtained from breaking up the hit grains and the energy imparted to powder particles while media hitting reach the equilibrium. This revised reading of the attenuation factor leads to the definition of the energy density transferred to powder at each impact ($E_0/V_0$) as the kinetic energy of impacting medium averaged across the active volume of powder nipped in a collision event. The energy balance between $E_0/V_0$ and $\gamma \cdot \Delta S_V$, the specific surface energy, defines the lowest physically attainable powder particle size, and conversely the largest specific surface.

This new formulation has been tested in three different settings: a laboratory for dry milling under strict supervision, a high energy ball mill for talc powder and an industrial pilot plant for wet ball milling an ultrafine silica slurry. The talc powder milling demonstrates that the procedure is also efficient when exfoliation is employed to reduce particle size instead of crushing.

Each of the situations under examination proves that $E_0/V_0$ and $\gamma$ are the primary factors determining the milling limit.

The process of material grinding can be understood as a chemical reaction triggered by the energy density during milling, which breaks atomic bonds creating new fracture surfaces. It is imperative to emphasize that the powder energy gain — a form of powder potential energy — is deemed to occur exclusively through particle fracturing. The suggested energy balance model cannot be used to properly understand processes like mechanical-alloying, milling over the material's fragile to ductile transition, or mechano-chemistry.

List of symbols

| Symbol | Description |
|---|---|
| $a$ | Radius of the circular hertzian surface of contact between ball and powder [m] |
| $A$ | Jar oscillation amplitude [m] |
| $d_0$ | Initial average particle size [m] |
| $d_{avg}$ | Average particle size [m] |
| $d_{lim}$ | Limiting particle size [m] |
| $d_r, d$ | Particle size [m] |
| $E_0$ | Impact energy [J] |
| $E_M$ | Energy per unit mass [J/kg] |
| $E_P$ | Elastic energy of a particle [J/m$^3$] |
| $E_V$ | Energy per unit volume [J/m$^3$] |
| $E_{ball}$ | Impact energy of a milling ball [J] |
| $f$ | Frequency [Hz] |
| $F$ | Applied load [N] |
| $G$ | Energy released by the body per unit fracture area [J/m$^2$] |
| $h$ | Height of the layer of powder [m] |
| $K_{Ritt}$ | Proportionality constant in the Rittinger's law [J/m] |
| $K_{SM}$ | Constant in the Rittinger's law expressed as specific surface per mass unit vs. time [m$^2$/kg/s] |
| $K_{SV}$ | Constant in the Rittinger's law expressed as specific surface per volume unit vs. time [m$^2$/m$^3$/s] |
| $m_{ball}$ | Mass of a milling ball [kg] |
| $N$ | Number of particles in the system [-] |
| $N_0$ | Starting number of particles in the system [-] |
| $r_{ball}$ | Radius of the milling ball [m] |
| $R_1, R_2$ | Principal radii of curvature of a surface [m] |
| $S_M$ | Surface per unit mass [m$^2$/kg] |
| $S_{M\infty}$ | Limiting surface per unit mass [m$^2$/kg] |
| $S_P$ | Surface of a particle [m$^2$] |
| $S_V$ | Surface per unit volume [m$^2$/m$^3$ = m$^{-1}$] |
| $S_{V0}$ | Starting surface per unit volume [m$^2$/m$^3$] |
| $S_{Vlim}$ | Limiting surface per unit volume [m$^2$/m$^3$] |
| $t$ | Time [s] |
| $v_{ball}$ | Speed of a milling ball [m/s] |
| $v_{jar}$ | Velocity of the oscillating jar [m/s] |
| $V_0$ | Active (Hertzian) volume of powder involved in the impact [m$^3$] |
| $V_P$ | Volume of a particle [m$^3$] |
| $x_{jar}$ | Position of the oscillating jar [m] |
| $x$ | Transformed fraction [-] |
| $Y$ | Elastic modulus [Pa] |
| $\beta_j$ | Factor expressing the increase of surface from one spherical particle to *j* spherical fragments [-] |
| $\gamma$ | Surface energy of the powder [J/m$^2$] |
| $\eta$ | Milling efficiency [-] |
| $\kappa$ | Surface curvature [m$^{-1}$] |
| $\nu$ | Poisson ratio [-] |
| $\rho$ | Density of the powder [kg/m$^3$] |
| $\sigma$ | Driving force for sintering [J/m$^3$] |

Acknowledgments
Dr. Eng. Daniela Di Martino (MBDA-Italy) is gratefully acknowledged for having kindly disclosed the data of the silica slurry preparation.
This work is dedicated to the memory of Roberto Ricci Bitti, our mentor.

Author contributions
SM: Conceptualization, Investigation, Writing – original draft, Writing – review & editing.
PEDN: Conceptualization, Investigation, Methodology, Formal Analysis, Writing – review & editing.

Funding
Not applicable

Data availability
A reasonable request for raw data can be addressed to the corresponding authors.

Declarations
*Conflict of interest*: On behalf of all authors, the corresponding authors state that there is no conflict of interest.

List of captions

Table 1. Calculated mass of powder trapped between the colliding ball and the jar wall, energy density and process efficiency optimized by non-linear least squares fitting (see Fig. 1) and additional calculated quantities. Stroke frequency is 20 Hz for all runs.

Table 2. Main results on talc milling with a high energy ball mill. All data are taken from the original paper of Kim et al. [11]. The last two columns show the ratio of the BET specific surface area measurements. The second to last column is the ratio among experimental values, the last one the calculated rate according to the present model. Note: the equivalent sphere diameter is $ESD = 6/(\rho\, S_M)$, where $\rho$=2800 kg/m$^3$.

Table 3. Values of energy density $E_0/V_0$ and efficiency $\eta$ optimized by non-linear fitting. The minimum particle size and the maximum specific surface are calculated from $E_0/V_0$ via Eq. (33). The last column represents the average particle size measured by Laser Diffraction (LD) and used as an industrial parameter for assessing the readiness of slurry.

Table 4. Comparison of the result from the analysis of Blanc et al.'s data and the outcomes from the wet milling of silica slurry. In curly brackets the ratio of the characteristic quantities: $E_0/V_0$, $d_{lim}$ and $S_{M\,lim}$, Blanc's data being the reference.

Fig. 1. Non-linear fit by eq. (35) of the experimental milling data by Blanc et al. [12] at a frequency of 20 Hz. The specific surface per unit mass is measured by laser diffraction (Mie's scattering).

Fig. 2. PSDs measured by laser diffraction on the 13 g sample milled at a frequency of 20 Hz by Blanc et al. [12] (open symbols). The experimental curve has been decomposed into the sum of two log-normal distributions (red and blue lines) to separate the components.

Fig. 3. Specific surface area as a function of the milling time for three different $ZrO_2$ ball sizes. The mill rotation speed is 2000 rpm.

Fig. 4. Particle size distribution (volume %) as a function of the milling time. Top: batch 107P; Bottom: batch 206P. Both batches have the same starting composition and have been milled under the same nominal conditions.

Fig. 5. Tumbler mill DEM simulation [45]. Number of collisions versus mean collisions energy. The dashed red rectangle points to the highest collision energy leading to the finest achievable particle dimension. The frequency of collision drops of two orders of magnitude towards the high energy collision side.

Fig. 6. Least squares fit to the experimental data sets. Left: batch 107P, right: batch 206P. The fitting results are shown in Table 3.